To be presented at *IEEE Symposium on Computational Intelligence for Financial Engineering (CIFEr);*
Canberra, Australia, 1st-4th December 2020

# Automated Creation of a High-Performing Algorithmic Trader via Deep Learning on Level-2 Limit Order Book Data

Aaron Wray
*Department of Computer Science*
*University of Bristol*
Bristol BS8 1UB, U.K.
wray1998@gmail.com

Matthew Meades
*Department of Computer Science*
*University of Bristol*
Bristol BS8 1UB, U.K.
mm16507@bristol.ac.uk

Dave Cliff
*Department of Computer Science*
*University of Bristol*
Bristol BS8 1UB, U.K.
csdtc@bristol.ac.uk

*Abstract*— **We present results demonstrating that an appropriately configured deep learning neural network (DLNN) can automatically learn to be a high-performing algorithmic trading system, operating purely from training-data inputs generated by passive observation of an existing successful trader *T*. That is, we can point our black-box DLNN system at trader *T* and successfully have it learn from *T*'s trading activity, such that it trades at least as well as *T*. Our system, called *DeepTrader*, takes inputs derived from Level-2 market data, i.e. the market's *Limit Order Book* (LOB) or *Ladder* for a tradeable asset. Unusually, DeepTrader makes no explicit prediction of future prices. Instead, we train it purely on input-output pairs where in each pair the input is a snapshot *S* of Level-2 LOB data taken at the time when *T* issued a quote *Q* (i.e. a bid or an ask order) to the market; and DeepTrader's desired output is to produce *Q* when it is shown *S*. That is, we train our DLNN by showing it the LOB data *S* that *T* saw at the time when *T* issued quote *Q*, and in doing so our system comes to behave like *T*, acting as an algorithmic trader issuing specific quotes in response to specific LOB conditions. We train DeepTrader on large numbers of these *S/Q* snapshot/quote pairs, and then test it in a variety of market scenarios, evaluating it against other algorithmic trading systems in the public-domain literature, including two that have repeatedly been shown to outperform human traders. Our results demonstrate that DeepTrader learns to match or outperform such existing algorithmic trading systems. We analyse the successful DeepTrader network to identify what features it is relying on, and which features can be ignored. We propose that our methods can in principle create an explainable copy of an arbitrary trader *T* via "black-box" deep learning methods.**

## I. Introduction

The motivation for our work is best explained by a brief sketch of where we hope to end up, a little two-paragraph story of a plausible near-future:

*Imagine a situation in which a highly skilled human trader operating in a major financial market has a device installed on her trading station, a small black box, with a single indicator lamp, that takes as input all data provided to the trader via her screen and audio/voice lines. The black box records a timestamped stream of all the market data that the trader is exposed to at her station, and also records a timestamped tape of all orders (quotes and cancellations) that she sends to the market: while it is doing this, the black box's indicator lamp glows orange, signaling that it is in Learning Mode.*

*After a while, maybe a few weeks, the indicator lamp on the black box switches from orange to green, signaling that it is now in Active Mode. At this point, the box starts to automatically and autonomously issue a stream of orders to the market, trading in the style of the human trader whose activity it has been monitoring. The box has learnt purely by observation of the inputs to the trader (market data and other information) and her outputs (various order-types) and its trading performance matches or exceeds that of the human trader. At this point the services of the human trader are no longer required.*

In the language of research psychologists, our approach sketched in this story is a *behaviorist* one: we are concerned only with the "sensory inputs" and "motor outputs" of the human trader, we do not care about (or, at least, we make no pre-commitment to) modelling her internal mental states, or her internal reasoning processes; we do not need to interview her in some "knowledge elicitation" process (cf. e.g. [6]) to find out what analysis she performs on the incoming data, what sequence of decisions leads her to issuing a particular order; we do not require our black box to internally compute a GARCH model, or even a MACD signal: all we ask is that when presented with a stream of specific market-data inputs, the outputs of our box is a stream of orders that lead to trading performance at least as good as the human trader that the box learned from.

In this paper, we demonstrate a proof-of-concept of such a system, called DeepTrader. We have not yet put it in a metal box with a single indicator lamp, but we've got the software working.

At the heart of DeepTrader is a Deep-Learning Neural Network (DLNN: see e.g. [9][12][22][28]), a form of machine learning (ML) that has in recent years been demonstrated to be very powerful in a wide range of application domains. DLNNs are instances of supervised learning, where training the ML system involves presenting it with a large training-set of 'target' input/output pairs: initially, when presented with a specific input, the output of the DLNN will be a long way from the target output, but an algorithm (typically based on the back-propagation of errors, or "backprop", introduced by [16]) adjusts the DLNN's internal parameters on the basis of the errors between the actual output for this specific input and the target output associated with that input, so that next time this input is presented, the difference between the actual and target outputs will hopefully be reduced. This process is iterated many times, often hundreds or thousands of cycles over training-sets involving many tens of thousands of target input/output pairs, and if all is well this leads to the errors reducing to acceptably small levels. Once the errors are small enough, the DLNN is hopefully not only producing close-to-target outputs for all of the input/output pairs in the training set, but it is also capable of generating appropriate outputs when presented with novel inputs that were not in the training set: i.e., it has generalized. For this reason, evaluating how well a DLNN has learned usually involves testing it post-training, on a test-set of input/output pairs that were not used in the training process.

In the fictional story we opened this section with, the input-output pairs in the training and test set would come from observing the human trader working her job in a real financial market: every time a significant event occurs in the market, an observable behavior of interest, that event or action is the desired output vector; and the associated input vector is some set of multivariate data that is believed to be necessary and sufficient for explaining the observable behavior of the trader – i.e. it is whatever data the trader is thought to have been exposed to and acting upon at the time the event occurred. In our work reported here, each input vector is calculated from a timestamped snapshot of a financial exchange's Limit Order Book (LOB) (also known as the Ladder in some trading circles), i.e. the array of currently active bids and offers at the exchange, represented as the prices at which there are limit orders currently resting, awaiting a counterparty, and the quantity (total size of orders) available at that price. The output vector, the action to be associated with each input, could be an order (a fresh quote, or a cancellation of a previous order) issued from the trader to the exchange, and/or it could be a trade executing on the exchange.

More generally, as a source of input-data for DeepTrader, we need a market environment, which we'll denote by M; and to generate the target outputs used in the training-set we need a training trader, which we'll denote by T. We think it arguable whether we actually need a test-set, as a standalone collection of fresh input-output pairs: in principle, once DeepTrader's DLNN training process has produced an acceptable drop in error-levels on the training set, then it could just be set to work on live trading in the market M – whether it makes a profit or a loss in that trading would then be the final arbiter of whether the learning was successful or not. Such an approach would suit risk-seeking developers who have sufficient funds available to take the financial hit of whatever losses an under-generalized DeepTrader makes before it is switched off: for the risk-averse, positive results from a test-set could provide useful reassurance of generalization before DeepTrader goes live.

Real professional human traders are typically very busy people who don't come cheap, and also there will most likely be some regulatory and internal-political hurdles to overcome if we did want to record the necessary amounts of data from a human trader, which would only serve to delay us. So, for our proof-of-concept reported here, we have instead used high-performing algorithmic trading systems (or "algos" for short) as our T, our training trader. Specifically, the algos that we use include two that have been repeatedly shown to outperform human traders in experiments that evaluated the trading performance of humans and algos under controlled laboratory conditions. These two "super-human" algos are known by the acronyms AA (for *Adaptive Aggressive*: [25][26]) and ZIP (for *Zero Intelligence Plus*: [4]). Given that these out-perform human traders, we reason that if DeepTrader's DLNN can be trained to match or exceed the trading behavior of these algorithms in the role of T, then the likelihood is that it will also do very well when we deploy the same methods albeit using data from a human T – this is a topic we return to in the discussion section at the end of this paper. Another advantage conferred by using algo traders as T at this stage is replicability: the source-code for the traders is in the public domain, and so anyone who wishes to replicate or extend the work we report here can readily do so.

Having identified a T to produce target outputs, we also need an M, a market environment to generate the inputs associated with each target output. Again, as this is a proof-of-concept study, instead of using data from a real financial market (with its associated nontrivial costs and licensing issues, and the difficulty of doing direct replication) we instead use a high-fidelity simulation of a contemporary electronic exchange. For the T in this study we use the long-established public-domain market-simulator BSE [2][5] as our source of input data. BSE is an open-source GitHub project written in Python, which was first made public in 2012, and provides a faithful detailed simulation of a financial exchange where a variety of public-domain automated trading algorithms interact via a Continuous Double Auction (CDA: the usual style of auction for all major financial exchanges, where buyers can issue bids at any time and sellers can issue asks/offers at any time) for a simulated asset: traders can issue a range of order-types (and cancellations), and BSE publishes a continuously-updated LOB to all market participants: it is timestamped snapshots of that LOB data that form the input data for training and testing the DLNN in DeepTrader. BSE includes a number of pre-defined algorithmic traders including AA and ZIP, so the Python source-code we used for our T traders can be found alongside the source-code we used for our M market, in the BSE GitHub repository [2].

The rest of this paper, much of which is drawn from [27], is structured as follows. In Section II we summarize the background to this work. Section III then describes our methods. In Section IV we present results which demonstrate that DeepTrader learns to outperform pre-existing algo traders in BSE, and matches or exceeds the trading ability of the two

"super-human" algorithms AA and ZIP. Section V (drawn from [15]) further analyses those results; we discuss our plans for further work in Section VI, and offer conclusions in Section VII.

## II. BACKGROUND

Our work reported here uses a public-domain simulator of a contemporary electronic financial exchange running a CDA with a LOB, so in that sense our work is very much about AI in present-day and future financial trading systems, but the roots of our work, and of the simulator we use, lie in academic economics research that commenced more than 50 years ago.

In 1962, Vernon Smith published an article in the prestigious *Journal of Political Economy* (JPE) on the experimental study of competitive market behaviour [19]. The article outlined a number of laboratory-style market simulation experiments where human subjects were given the job of trading in a simple open-outcry CDA where an arbitrary asset was traded, while the experimenters looked on and took down their observations. The supply and demand curves used in these experiments were realistic, but were predetermined by Smith, who allocated each trader a private limit price: the price that a buyer cannot pay more than, or the price that a seller cannot sell below. Different buyers might be given different limit prices, and the array or schedule of limit prices would determine the shape of the demand curve in the experimental market; ditto for the schedule of sellers' limit prices and the resultant market supply curve. In this sense, Smith's experimental subjects were like sales traders in a brokerage or bank, running customer orders: some external factor sets a limit price, and the trader's job is to do their best to buy or sell within that limit. If a buyer can get a deal for less than her limit price, the difference is a saving; if a seller can get a deal for more than her limit price, that's profit. Economists use 'utility' or 'surplus' to refer to both the buyer's difference and the seller's difference, but as we're focused on applications in finance we'll use 'profit' for both.

The experiments run by Smith demonstrated a rapid convergence of a market to its theoretical equilibrium price (the price where the quantity of goods supplied is equal to the quantity of goods demanded, where the supply curve intersects the demand curve) in a CDA, even with a small number of traders. This was measured by using Smith's 'a' metric, a measure of how well transactions in the market converge on the equilibrium price. In 2002, Smith received the Nobel Prize in Economics for his work establishing the field of experimental economics, and variations of his experiments have become *de facto* standards for test and comparison of trading algorithms.

Winding forward roughly 30 years, in 1990 a competition was hosted at the Santa Fe Institute for designing the most profitable automated trading agent on a CDA [17]. Thirty contestants competed for cash prize incentives totaling $10,000. The prize money won by each contestant was in proportion to the profit that their agent received in a series of different market environments. The highest ranked algorithm, designed by Todd Kaplan, was a simple agent that would hide in the background and hold off from posting a bid/ask price whilst letting other traders engage in bidding negotiations. Once the bid/ask spread was within an adequate range, Kaplan's agent would enter and "steal the deal". Aptly, Kaplan's program was named *Sniper*. If the market session was about to end, Sniper was programmed to rush to make a deal rather than not make one at all.

Subsequent to this, in 1993 Gode & Sunder published a JPE paper investigating the intelligence of automated traders and their efficacy within markets [21]. They developed two automated trading agents for their experiments, the Zero-Intelligence Unconstrained (ZIU) and the Zero-Intelligence Constrained (ZIC). The ZIU trader generates completely random quote prices, whereas the ZIC trader quotes random prices from a distribution bounded by the trader's given limit price, so the ZIC's are constrained to not enter loss-making deals. Gode & Sunder's series of experiments were performed in a similar style and spirit to Smith's: they ran some human-trader experiments to establish baseline data, and then ran very similar experiment with markets populated only by ZIU traders, and then only by ZIC. In each market they recorded three key metrics: allocative efficiency; single agent efficiency; and profit dispersion. The allocative efficiency is a measure of the efficiency of the market. It is the total profit earned by all traders divided by the maximum possible profit, expressed as a percentage. Gode & Sunder's key result was that the allocative efficiency of ZIC markets was statistically indistinguishable from that of human markets, and yet allocative efficiency had previously been thought to be the marker for intelligent trading activity. Ever since, ZICs are used as a de facto standard benchmark for a lower-bound on automated traders.

Extending the work of Gode & Sunder, in 1997 Cliff [4] identified that there were certain market conditions where ZIC traders would fail to exhibit human-like market dynamics. This finding led Cliff to create an automated trading agent with some elementary added AI, one of the first adaptive automated traders, called Zero Intelligence Plus (ZIP). The ZIP trader calculates its own profit margin which, along with its given limit price, it uses to calculate its bid or ask price. The profit margin is determined by a simple machine-learning rule and is adjusted depending on the conditions of the market. If trades are occurring above the calculated price, the profit margin is increased/decreased depending on whether the trader is a buyer/seller.

At roughly the same time as Cliff was publishing ZIP, in 1998 Gjerstadt & Dickhaut co-authored a paper that approached the sales trader problem from a new perspective [16]. They developed a price formation strategy in a CDA that analyzed recent market activity to form a belief function. The frequencies of bids, asks, accepted bid and accepted asks, from a set number of the most recent trades were used to estimate the belief or probability that an ask or bid would be accepted at any particular price. With this trading strategy, which came to be widely referred to as the GD strategy, the function selects an ask/offer price that would maximize a trader's expected gain based on the data. The strategy produced efficient allocations and was found to achieve competitive equilibrium within markets.

Then in 2001 a team of IBM researchers modified GD by interpolating the belief function to smooth the function for prices that did not occur in the selected number of recent trades, and they named the new trading agent MGD (Modified GD) and published results in a paper at the prestigious International Joint Conference on AI (IJCAI) that generated worldwide media coverage [7]: the IBM team was the first to explore the direct interaction between automated trading agents and human traders in a methodical manner, using LOB-based CDA markets that were close to ones implemented in financial exchanges across the world, where the traders in the market were a mix of human traders and automated algorithmic traders (specifically: IBM's MGD, Kaplan's Sniper, Gode & Sunder's ZIC, and Cliff's ZIP). Famously, the IBM team demonstrated that MGD and ZIP could consistently outperform human traders in these realistic market scenarios – that is, MGD and ZIP are `super-human' traders. And the rest, as they say, is history: the IBM work got the attention of many investment banks and fund-managmenet companies, and in following years the world of finance started to see ever increasing levels of automation, with more and more human traders replaced by machines.

Academic and industrial R&D continued after the landmark IBM study, and two significant subsequent developments were the extension of MGD into GDX, and a new ZIP-related trading algorithm called AA. Details of GDX were published in 2002 by Tesauro & Bredin [23]: GDX exploits dynamic programming to learn functions that better incorporate long term reward, and at the time it was published IBM claimed it as the world's best-performing public-domain trading strategy. Details of AA were published by Vytelingum in his PhD thesis [25] and subsequent article in the prestigious *Artificial Intelligence* journal [26]. The key element of AA is *aggressiveness*: a more aggressive trader places a bid/ask that is more likely to be accepted, while a less aggressive trader will aim to seek a larger gain. This trading strategy estimates the market's equilibrium price by using a weighted moving average and estimates the volatility of the market by using Smith's $\alpha$ metric.

Inspired by the IBM experiments pitting human traders against robot traders, a decade later in 2011 De Luca and Cliff ran a series of experiments, reported at IJCAI in [8], which suggested that AA dominates all known trading strategies and also outperforms humans, making AA the third trading strategy to be demonstrated as super-human. However, recently Snashall and Cliff [20] performed a brute-force exhaustive search of all possible ratios or permutations of different trading strategies for markets populated by a specific number of traders, consisting of over 1,000,000 market sessions, in order to show that AA doesn't always outperform GDX or ZIP: there are some circumstances in which AA can be dominated by GDX, or by ZIP.

While AA, GD, GDX, MGD, and ZIP were all early instances of AI in finance, in virtue of their use of machine learning (ML) to adapt to circumstances and outperform human traders, they all used relatively simple and traditional forms of ML. In the past decade there has been an explosion of interest in Deep Learning, the field that concentrates on solving complex problems through the use of "deep" (many-layered) neural networks, i.e. DLNNs.

It is commonplace to implement recurrent DLNNs for time series forecasting and a vast amount of research has been completed in this area particularly in spot markets where traders attempt to predict the price of a resource in the future. Predictions are often made to assist in generating a signal on whether a trader should buy, hold or sell the resource that they are trading. Although this project employs a DLNN, there is a clear distinction on how it is being used. Rather than being used to predict a future price, this DLNN will be applied to the sales trader problem directly: a DLNN (specifically, a Long Short Term Memory, or LSTM DLNN: see [12]) is created that receives a limit price from customer orders, considers the conditions in the market by extracting information from the LOB, and finally given all of this information produces a price to quote in the next order, a desired price to transact at.

To the best of our knowledge, there are only two pieces of work that are closely related enough to discuss here. The first is DeepLOB [28] which uses a form of DLNN traditionally used in image processing, to capture the spatial structure of a LOB, coupled with an additional recurrent DLNN that incorporates information gathered over long periods of time. The second is by Le Calvez and Cliff [14] which demonstrates preliminary results from the use of a DLNN to successfully replicate the behavior of a ZIP trader, but which used only the best bid and ask prices. As Sirignano and Cont [18] have recently and elegantly demonstrated, deeper (Level-2) LOB data can be highly informative about short-term market trends, so a natural question to explore given Sirignano and Cont's result is: can we extend the methods reported by Le Calvez and Cliff to instead use Level-2 LOB data? That is what we explore in this paper.

### III. METHODS

Comparing the performance of trading strategies is not a straightforward task. As previously mentioned, the performance of a strategy is reliant upon the other traders within the market and in real-world financial markets, it is implausible to know what algorithms other traders are using, as this information is confidential. Traders tend not to disclose their strategies in order to remain profitable, for obvious reasons. Nevertheless, there are well-established experiment-methods which can be used to compare trading agents. IBM's Tesauro and Das [24] present three separate experiment designs for comparing trading agents, two of which will be used here: in one-in-many tests (OMTs), one trader is using a different strategy to the all rest -- this test is used to explore a trading strategy's vulnerability to invasion and defection at the population level; and in balanced-group tests (BGTs) buyers and sellers are split evenly across two types of strategy, and for every trader using strategy A is matched with a trader using strategy B, with the matched-pair ech being given the same limit price. BGTs are generally considered to be the fairest way to directly compare two strategies.

BSE was used to generate and collect all of the data required to train the LSTM network for DeepTrader and then test its performance against existing trading strategies. BSE allows control of the supply and demand schedules for a market session: we specified a range of schedules with varying shapes to both the supply and the demand curves, to generate data from a wide range of market conditions.

BSE produces a rich flow of data throughout a market session, including a record of the profit accumulated by each trader: when we present our results in Section 4, we focus on average profit per trader (APPT) because this is metric is reassuringly close to the profit and loss (P&L) figure that real-world traders (humans or machines) are judged by.

The LOB maintained by BSE is updated and published to all traders in the market whenever a new limit order is added to it,

whenever a market order executes, or whenever an order is cancelled (thereby taking liquidity off the LOB). The published LOB is represented within BSE by a data-structure made up of an order-book for bids, and an order-book for asks. Each of these two order-books contains a list of the prices at which orders are currently resting on the book, and the quantity/size available at each such price. From this LOB data, it is possible to calculate various derived values such as the bid-ask spread, the mid-price, and so on. BSE also publishes a 'tape' showing a time-ordered list of timestamped market events such as orders being filled (i.e., transactions being consummated) or being cancelled. The clock in BSE is usually set to zero at the start of a market session, so the time t shows how much time has elapsed since the current market session began.

DeepTrader takes as input 14 numeric values that are either directly available on BSE's LOB or tape outputs, or directly derivable from them: these 14 values make up the 'snapshot' that is fed as input to DeepTrader's LSTM network for each trade that occurred within a market session. The 14 values are as follows (the +/− prefixes on input values are used in Section V):

1. − The time $t$ the trade took place.
2. + Flag: did this trade hit the LOB's best bid or a lift the best ask?
3. + The price of the customer order that triggered the quote that initiated this trade.
4. + The LOB's bid-ask spread at time $t$.
5. + The LOB's midprice at time $t$.
6. + The LOB's microprice at time $t$.
7. + The best (highest) bid-price on the LOB at time $t$.
8. − The best (lowest) ask-price on the LOB at time $t$.
9. − The time elapsed since the previous trade.
10. − The LOB imbalance at time $t$.
11. − The total quantity of all quotes on the LOB at time $t$.
12. − An estimate $P^*$ of the competitive equilibrium price at time $t$, using the method reported in [25][26].
13. − Smith's $\alpha$ metric [19], calculated from $P^*$ at time $t$.
14. The price of the trade.

The first 13 items in the list are the inputs to the network: if any of them is undefined at time $t$ then zero is used. Item 14 is the output (target) variable that the network is training toward: this is the price that DeepTrader will quote for an order. With respect to Item 3 on this list, it is important to note that when DeepTrader is trading live in the market, it only has access to the limit-prices of its own customer orders.

Each of these 13 input variates can have values within differing ranges. As a single input consists of 13 different features and the contribution of one feature depends on its variability relative to other features within the input. If for example, one feature has a range of 0 to 1, while another feature has a range of 0 to 1,000, the second feature will have a much larger effect on the output. Additionally, values in a more limited range (e.g. 0 to 1) will result in faster learning. Therefore, when training a multivariate neural network such as in DeepTrader, it is common practice to normalize all features in the training dataset such that all values are within the same scale. We used min-max normalization: for further details see [25].

The BSE GitHub repository [2] includes source code for seven different trading strategies, four of which (AA, GDX, ZIC, & ZIP) have already been introduced. The remaining three are SNPR, a trader directly inspired by (but not identical to) Kaplan's Sniper; GVWY, a "giveway" trader that simply quotes at its own limit price, giving away all potential profit; and SHVR, a "shaver" trader whose strategy is simply always to undercut the current best ask by one penny, and/or to always beat the current best bid by one penny – this strategy is intended as a minimal model of a pesky high-frequency trader.

To create a large dataset to train the model, many market-session configurations were devised where the proportions and types of traders were varied. Each market session had 80 traders (40 buyers and 40 sellers). Additionally, each market session involved four different trading strategies. For each trading strategy, the number of buyers and sellers was always the same but there were five different proportion-groups of traders used. These proportion-group were: (20, 10, 5, 5), (10, 10, 10, 10), (15, 10, 10, 5), (15, 15, 5, 5), and (25, 5, 5, 5). Each number within a group denotes the number of buyers and sellers for a specific trading strategy within a market session. For example, the (20, 10, 5, 5) proportion group, indicates that there were 20 buyers and sellers of trading strategy 1; 10 buyers and sellers of trading strategy 2; 5 buyers and sellers of trading strategy 3; and 5 buyers and sellers of trading strategy 4 within this group.

Given that there are four trading strategies in each session selected from a total pool of 7 available strategies, there is a total of 35 different combinations (i.e. 7 combined 4).

Furthermore, there are 35 different permutations for each of the proportion groups listed. This led to a total of 1225 (=35x35) different market configurations where the proportions and types of traders were varied. Each market configuration was executed 32 times with different random-number sequences for additional variability giving a total of 39,200 different market sessions that were run to create the training data for DeepTrader.

Each individual market session takes approximately 30 seconds to complete, so running all 39,200 on a single computer would take approximately 13.5 days. For this reason, the decision was made to use Amazon's *Elastic Compute Cloud* (EC2) service to parallelize data generation and collection processes amongst 32 virtual machines (VMs). The Python library *Boto3* v.1.13.3 [10] was used to create, manage and terminate the VMs. Work was automatically split amongst the VMs by making every VM run each market configuration once and via a separate custom utility, created for this project.

Typically, neural networks have *training*, *validation* and *test* datasets. The training set is used to train the model, it is the data that a neural network learns from; whilst the validation dataset is used for tuning a model's hyperparameters, and the test dataset is used to evaluate a model's final performance. For this project, as the performance of the neural network is determined by how well it trades during a market session, the test data is generated dynamically as the trader interacts with the simulated market.

The LSTM network created consists of three distinct hidden layers. The first hidden layer is an LSTM layer containing 10 neurons. The final two hidden layers are both fully connected layers containing 5 and 3 neurons respectively. Each hidden layer uses the Rectified Linear Unit (Relu) as an activation function: further details are given in [27].

The training process is limited by the size of memory on the machine used to train the network: the training dataset was so large that using all data points at once is not practicable because it exceeds the memory limits of conventional commodity servers, and we did not have a national-scale supercomputer readily available. Therefore, was training was executed in batches. Each batch consisted of 16,384 data points and the Adam optimizer [13] is used to train the network. As is common in DLNN applications, the network's learning rates require careful selection, and Adam uses an adaptive learning rate method that calculates different learning rates based on the weights in the network, with the intention of finding a workable tradeoff between overfitting (if the learning rate is set too high) and long processing times (if it is set too low).

The function that was used to calculate the error (loss) within the network was the mean squared error (MSE), as described in more detail in [25]. An epoch in training is the network being presented with each data-point within the training dataset once. We trained DeepTrader's LSTM network for 20 epochs, and the error measure typically fell rapidly in the first 10 epochs and was thereafter asymptotic approaching a very low value for the remainder of the training process. So, in total, DeepTrader would be trained via exposure to LOB data from 20 x 39,200 = 784,000 individual market sessions, and each of those sessions would typically involve roughly 20 LOB snapshots, so the total number of snapshots used in training was around 15 million.

## IV. RESULTS

Figures 1 to 4 show box-plots summarizing results from our experiments. Each experiment involves n=100 independent and identically distributed trials in the particular market, with a different sequence of random numbers generated for each trial. In all these figures, the vertical axis is average profit per trader (APPT) and the box is plotted such that distance between the upper and lower edges is twice the inter-quartile range (IQR); the horizontal line within the box is the median, and any data-points that are more than 1.5 times the IQR from the upper or lower quartiles are regarded as outliers and plotted individually. Figures 1 and 2 show results from the balanced group tests (BGTs), while Figures 3 and 4 show results from the one-in-many tests (OMTs). As is described in detail in Chapter 4 of [25], as a test of the significance of the differences observed between the APPT for DeepTrader and the APPT for whatever pre-existing algorithm it is being tested against, we calculated 90% confidence intervals (CIs) around the mean and judged the difference in distributions to be significant if the CIs of the two trading strategies were non-overlapping.

Fig. 1a shows BGT comparison of APPT scores between DeepTrader and ZIP, and Fig. 1b shows BGT comparison of APPT scores between DeepTrader and AA. Comparison of 90% confidence intervals around the mean APPT scores for the two trading strategies in each graph indicates that in this case there is no significant difference between ZIP and DeepTrader, but AA does significantly outperform DeepTrader. However, as we will see in Fig. 3, when AA is pitted against DeepTrader in one-in-many tests, AA is outperformed by DeeTrader: we discuss these AA results later in this section.

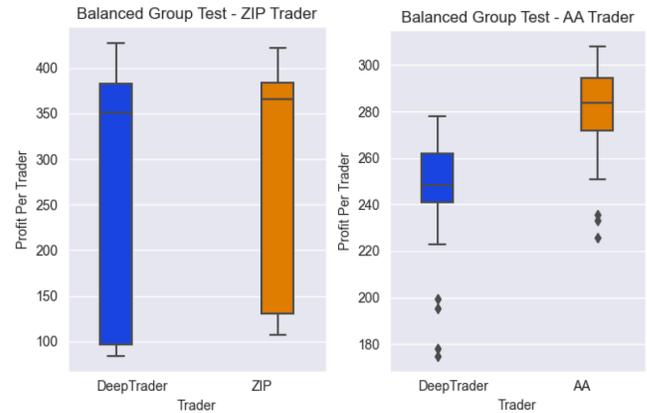

Fig. 1. Box-plots showing average profit per trader (APPT) from balanced-group tests (BGTs) for DeepTrader vs ZIP algorithmic traders (left: Fig1a) and AA algorithmic traders (right: Fig1b).

Fig. 2a shows BGT comparison of APPT scores between DeepTrader and GDX, and Fig. 2b shows BGT comparison of APPT scores between DeepTrader and ZIC. Comparison of 90% confidence intervals around the mean APPT scores for the two trading strategies in each graph indicates that in this case DeepTrader significantly outperforms GDX, and ZIC too.

Fig. 3a shows OMT comparison of APPT scores between DeepTrader and ZIP, and Fig. 3b shows OMT comparison of APPT scores between DeepTrader and AA. Comparison of 90% confidence intervals around the mean APPT scores for the two trading strategies in each graph indicates that DeepTrader significantly outperforms both ZIP and AA, although from visual inspection of the graphs it is also obvious that DeepTrader has much more variability of response than either ZIP or AA.

Fig. 4a shows OMT comparison of APPT scores between DeepTrader and GDX, and Fig. 4b shows OMT comparison of APPT scores between DeepTrader and ZIC. Comparison of 90% confidence intervals around the mean APPT scores for the two trading strategies in each graph indicates that DeepTrader significantly outperforms both GDX and ZIC, although again from visual inspection of the graphs it is also obvious that DeepTrader has much more variability of response than either GDX or SHVR.

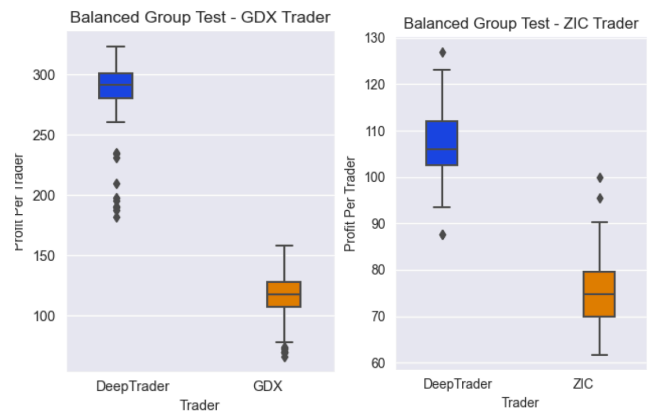

Fig. 2. Box-plots showing APPT from BGTs for DeepTrader vs GDX algorithmic traders (left: Fig.2a) and ZIC algorithmic traders (right: Fig.2b).

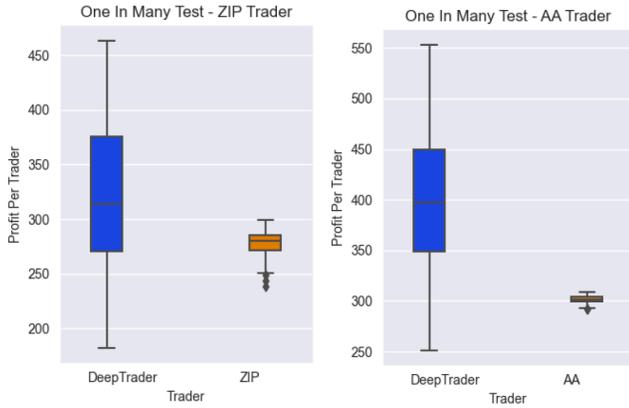

Fig. 3. Box-plots showing APPT from one-in-many tests (OMTs) for DeepTrader vs ZIP traders (left: Fig3a) and vs AA traders (right: Fig3b).

The results presented here demonstrate that DeepTrader achieves what we set out to do: when trained on a series of orders issued by a trader T, where each order is associated with a snapshot of the Level2 market data available to T at the instant that the order was issued, the DLNN in can be trained such that DeepTrader learns a mapping from the inputs (Level 2 market data inputs to DeepTrader) to outputs (quotes issued by DeepTrader) that result in superior trading performance when the final trained DeepTrader system is evaluated by allowing it to live-trade in the market environment M that the original trader T was operating in.

DeepTrader equals or outperforms the following trading algorithms in both the balanced group tests and the one-in-many tests: GDX, SHVR, SNPR, ZIC, and ZIP. Space limitations prevent us from including here further results, presented in [27], which show DeepTrader similarly learning to equal or outperform the rest of BSE's built-in algorithmic traders, i.e. GVWY, SHVR, and SNPR, thereby taking the total number of trading algos that DeepTrader outperforms to six. The most notable aspect of DeepTrader learning to trade at least as well as, or better than, this list of six different algorithms is that it includes ZIP, one of the two "super-human" algo traders for which code is already available in BSE.

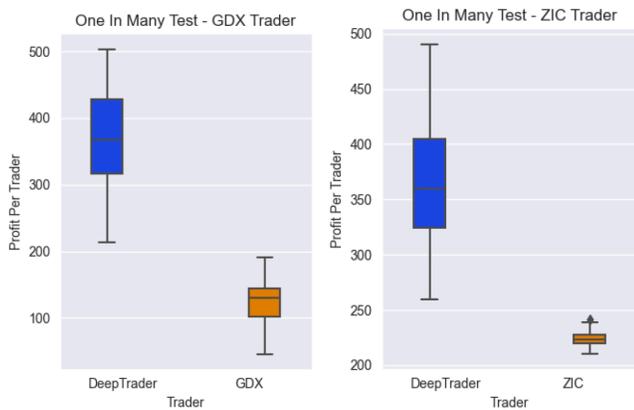

Fig. 4. Box-plots showing APPT from OMTs for DeepTrader vs GDX (left: Fig4a) and vs ZIC (right: Fig4b).

The results for DeepTrader when learning from (and then pitted against) the final algo studied here, Vytelingum's AA, the other super-human algo trader in BSE is somewhat less clear-cut: in the balanced group test, AA gets the better of DeepTrader; but in the one-in-many tests, DeepTrader roundly outperforms AA. As these two sets of tests result in a 1-1 tie, it seems fair to call it a draw.

So, in summary the results presented here (which are expanded upon in [27] and [15]) collectively show DeepTrader having six clear wins and one draw. While seven straight wins would naturally be preferable, these results nevertheless clearly demonstrate that the approach we have developed here has merit, and warrants further exploration.

In particular, having successfully used deep learning to create such a successful trader, the success provokes a natural question: how does DeepTrader work? That is, in what way does it use the 13 input features when trading? This is a question that we start to answer in the next section.

## V. ANALYSIS

Thus far our focus has been on ablation studies: that is, removing, disabling or masking specific aspects of the DeepTrader network and then noting the effects of that ablation. In general terms, we perform a sequence of ablation studies and the results from such a sequence can prompt us to hypothesise about how we could increase the efficiency of DeepTrader; we then test such hypotheses by studying the market performance of edited versions of DeepTrader, ones that have been altered to reflect our hypotheses – and this process can be iterated in an attempt at reducing the DeepTrader network to a minimally complex version that retains the ability to produce the desired level of trading behavior.

As is described in more detail in [15] we ran a set of 13 experiments where in each experiment one of the 13 input features was "ablated" by having the values in that column of the training data-set randomly shuffled (so the frequency-distribution of values in that column was unaltered, but any correlation between the value in that column and the other 12 features was very heavily disrupted); this gave us 13 sets of results where we could record the performance hit, the increase in error rate, when a specific one of the 13 features was ablated. A relatively large performance hit for a specific feature is an indication that DeepTrader's good trading behavior is indeed to reliant on that feature, whereas if the performance hit was sufficiently low then we considered that as an indication that the feature in question was, to a first approximation, irrelevant (or, at least, of sufficiently limited significance that it could be safely ignored) – that is, a low performance hit for ablating a feature led us to hypothesize that DeepTrader could operate successfully without access to that feature as an input; to check this, we re-ran our experiments with the feature (and its associated neurons within DeepTrader) absent, to check that the behavior of the resultant trader was consistent with our belief.

Via this method, we eliminated 7 of the 13 features (those prefixed with a – character in this enumerated list of 13 features in Section III), leaving a 6-input DeepTrader network which, when re-trained and tested, proved to show a slim improvement in its validation loss result: i.e. eliminating the seven features identified by ablation studies and using only the remaining six did not cause any loss of performance, and actually gave a very slight improvement. Thus we conclude this paper with the observation that only the six features prefixed with a + symbol in the enumerated list in Section III are required to trade as well as the 'super-human' trading agents ZIP and GDX.

## VI. FURTHER WORK

Although at the start of this paper we characterized our approach to the use of learning in DeepTrader as *behaviorist*, because we concentrate only on the observable inputs and resultant trading behavior of the system, our next phase of work will be devoted to further analysing the internal mechanisms that make a trained DeepTrader so successful. In particular, we will investigate the extent to which each of the key inputs identified in Section V contribute to the behavior of DeepTrader in a variety of market conditions: it is possible that some of those inputs play a much more significant role than others, and it is possible that which inputs are most significant varies across all market conditions: we will report on our findings in this respect in a future publication; there is much to explore. And, given the story that we opened this paper with, an obvious next step is to attempt to repeat the methods used here, with the source data coming from human traders rather than from trading agents.

## VII. DISCUSSION AND CONCLUSIONS

We have explored here the problem of using machine learning to automatically create high-performance algorithmic traders that are fit to operate profitably in a contemporary financial exchange based (as are all current major electronic exchanges) on a CDA process mediated by a continuously updated limit order book showing Level 2 market data. Our approach to this problem is behaviorist, in the sense that we seek to use machine learning to replicate or exceed the trading behavior of an existing high-performance trader, and we do this purely by specifying a set of desired outputs for particular inputs: we have made no commitment to any particular approach being incorporated within the trader's internal processing that maps from externally observable inputs to outputs; instead we treat DeepTrader as an opaque black-box.

The novel results presented in this paper demonstrate for the first time this approach being used successfully against a range of pre-existing algorithms, including both AA and ZIP which had previously been shown to outperform human traders. Given that AA and ZIP are already known to exceed the capabilities of human traders in LOB-based CDA markets, it seems plausible to conjecture that the methods used here could in principle be extended to operate on training data that comes from observation of a human trader rather than an algorithmic trader. The basic approach, of associating snapshots of the LOB with orders issued by the trader, should work independently of whether the trader issuing the order is a person or a machine. And, in that sense, the little story that we started this paper with may not be fiction for much longer.